\documentclass[dvips,, 12pt]{article}
\usepackage{latexsym, epsfig, amsmath, graphicx, amssymb}
\usepackage{mathrsfs, amsfonts, graphics}

\begin{document}

\begin{center}
\textbf{\Large On the Existence of New Conservation
Laws for the Spaces of Different Curvatures}\\[1.2cm]
Tooba Feroze \\[3ex]
Centre for Advanced Mathematics and Physics, National University of
Sciences and Technology, Campus of EME College, Peshawar Road, Rawalpindi, Pakistan \\[0pt]
E-mail: tferoze@camp.edu.pk \\[0pt]
\end{center}

\bigskip

\begin{quotation}
\emph{\textbf{Abstract}\newline It is known that corresponding to
each isometry there exist a conserved quantity. It is also known
that the Lagrangian of the line element of a space is conserved.
Here we investigate the possibility of the existence of ``new"
conserved quantities, i.e. other than the Lagrangian and associated
with the isometries, for spaces of different curvatures. It is found
that there exist new conserved quantities only for the spaces of
zero curvature or having a section of zero curvature.}
\bigskip
\end{quotation}

\textbf{Introduction}

Sophus Lie investigated the role of transformation theory in
classical integration methods. He studied continuous transformations
that depend on parameter(s)
\begin{equation}
s^{\ast }=s^{\ast }\left( s,x^{i};\varepsilon \right) ,\;\;x^{i\ast
}=x^{i\ast }\left( s,x^{i};\varepsilon \right),   \label{1}
\end{equation}
and satisfy the axioms of a group and, hence, form a group of point
transformations. Then he, with the help of these transformations
(called the symmetry transformations or symmetries), provided
different integration strategies for differential equations. An
important class of these symmetries which leave the action principle
invariant is that of the Noether symmetries, i.e. $\mathbf{X}$ is
defined as a Noether symmetry if
\begin{equation}
\mathbf{X}\int L\left( s,x^{i},\dot{x}^{i}\right) ds=0;i=1,2,...,n
\label{2}
\end{equation}
where $L$ is the Lagrangian, $x^{i}\mathbf{=}\left(
x^{1},x^{2},x^{3},...x^{n}\right) $\textbf{\ }is a point in the $n-$
dimensional underlying manifold, $\dot{x}^{i}$\textbf{\ }is the
derivative of $x^{i}$ with respect to $s$ and
\begin{equation}
\mathbf{X}=\xi \left( s,x^{i}\right) \frac{\partial }{\partial
s}+\eta ^{a}\left( s,x^{i}\right) \frac{\partial }{\partial
x^{a}}+\dot{\eta} ^{a}\left( s,x^{i},\dot{x}^{i}\right)
\frac{\partial }{\partial \dot{x}^{a}} ,\;a,i=1,2,...,n  \label{3}
\end{equation}
where
\[
\xi \left( s,x^{i}\right) =\left. \frac{\partial s^{\ast }}{\partial
\varepsilon }\right| _{\varepsilon =0},\;\;\eta ^{a}\left(
s,x^{i}\right) =\left. \frac{\partial x^{a\ast }}{\partial
\varepsilon }\right| _{\varepsilon =0},\;\;\dot{\eta}^{a}\left(
s,x^{i},\dot{x}^{i}\right) =\frac{ d\eta
^{a}}{ds}-\dot{x}^{a}\frac{d\xi }{ds}.
\]
For some gauge function $A\left( s,x^{i}\right) $ eq.$\left(
\ref{2}\right) $ can equivalently be given as
\begin{equation}
\mathbf{X}L+L\frac{d}{ds}\xi =\frac{d}{ds}A,  \label{4}
\end{equation}
where
\[
\frac{d}{ds}=\frac{\partial }{\partial s}+\dot{x}^{a}\frac{\partial
}{
\partial x^{a}}.
\]
For each Noether symmetry one can reduce the equation of motion
(Euler-Lagrange equation) in order by two [1]. Also, corresponding
to each symmetry a conservation law is associated $\left[ 2\right]
$, and a conserved quantity, $T$, is obtained as [1]
\begin{equation}
T= \xi(\dot{x}^{a}L_{\dot{x}^{a}}-L)-\eta^{a}L_{\dot{x}^{a}}+A\left(
s,x^{i}\right).   \label{4a}
\end{equation}

In Differential Geometry the usual Lagrangian for the line element
\begin{equation}
ds^{2}=g_{ab}\left( x^{c}\right) dx^{a}dx^{b},\;\;a,b,c=1,2,...,n  \label{5}
\end{equation}
is
\begin{equation}
L\left( s,x^{a},\dot{x}^{a}\right)=1=g_{ab}\left( x^{c}\right)
\dot{x}^{a}\dot{x}^{b}.
\end{equation}

One may find isometries (functions that preserve the distance
between two points on a manifold) for the metric given by eq.$\left(
\ref{5} \right) $. In the context of isometries a lot of work has
been done $\left[ 3,4\right] $. One can associate geometrically
conserved quantities to isometries. One may also find the Noether
symmetries for the action of the line element. Recently, Bokhari
\emph{et al} find the Noether symmetries associated with the usual
Lagrangian of differnt line elements $\left[ 5\right] $. These
Noether symmetries may provide \emph{new} conserved quantities. The
aim of this paper is to generalize the idea and obtain general
result for the existence of new conserved quantities.

\textbf{Existence of New Conservation Laws}

 Before discussing the existence of conservation laws we first prove the
following

\textit{\textbf{Theorem 1:}
\begin{equation}
\mathbf{X}=s\frac{\partial }{\partial s}\\
\end{equation}\
is not a Noether symmetry generator associated with the Lagrangian
for the line element.}

\textit{\textbf{Proof:}} For $\mathbf{X}$ to be a Noether symmetry
it must satisfy eq.$\left( \ref{4} \right) $. But then
\begin{equation} \frac{d}{ds}A=-L\left( s,x^{a},\dot{x}^{a}\right),
\end{equation}\
or \begin{equation} \left( \frac{\partial }{\partial
s}+\dot{x}^{i}\frac{\partial }{\partial x^{i}}\right) A\left(
s,x^{c}\right)=-g_{ab}\left( x^{c}\right) \dot{x}^{a}\dot{x}^{b},
\end{equation}\
which is not possible as the left hand side can not be quadratic in
velocities. So it is not a Noether symmetry associated with the
Lagrangian for the line element.

Now, a result on the existence of new conservation law(s) is given
in the form of the following

\textit{\textbf{Theorem 2: (i)} There exist new conservation laws
for the spaces of zero curvature or for the spaces having a section
of zero curvature; (ii) There does not exist new conservation laws
for a space of non-zero curvature and not having a section of zero
curvature.}

\textit\textbf{Proof:} For the proof of this theorem two already
known results are required. The first result is that the algebra of
the Noether symmetries form a subalgebra of the symmetries of the
differential equations (geodesic equations or equations of motion)
$\left[ 1\right] $. The second result is that for a space of
non-zero curvature with isometry algebra $h$, the symmetry algebra
of the geodesic equations is $h\oplus d_{2}$ provided that there is
no section of zero curvature; if there is an m-dimensional section
of zero curvature and the isometry algebra for the orthogonal space
is $h_{1}$, then the symmetry algebra of the geodesic equations is
$h_{1}\oplus sl(m+2)$ $\left[ 6\right] $, where $d_{2}$ is a
$2-$dimensional dilation algebra with basis
\begin{equation}
\mathbf{X}_{0}=\frac{\partial }{\partial s},\;\mathbf{X}_{1}=s\frac{\partial
}{\partial s}.\;  \label{6}
\end{equation}
(i) For the n-dimensional spaces of zero curvature the isometry
algebra is $so(n)\otimes_{s}\mathbb{R}^{n}$, where $\otimes_{s}$ is
the semi direct product, and the symmetries of the geodesic
equations form $sl(n+2,\mathbb{R})$. Let two sets of symmetries
forming the isometry algebra and algebra of the symmetries of the
geodesic equations be S and I. Since the Noether symmetries form a
subalgebra of the symmetries of the geodesic equations, therefore,
for the spaces with zero curvature the new conservation laws are
associated with the Noether symmetries spanned by the space S-I. The
conserved quantity associated with $\mathbf{X}_{0}$, which is also
an element of $sl(n+2,\mathbb{R})$, is the Lagrangian. So, finally,
the new conservation laws can be obtained from the space
$S-I-\mathbf{X}_{0}$.

Similarly, for the spaces of non-zero curvature with an
m-dimensional section of zero curvature admitting the isometry
algebra $h_{1}\oplus[so(n)\otimes_{s}\mathbb{R}^{n}]$ and the
symmetry algebra of the geodesic equations  $h_{1}\oplus sl(m+2)$,
the space spanned by $S-I-\mathbf{X}_{0}$ may give new conservation
laws.

(ii) For the spaces of non-zero curvature there are only two
additional symmetries of the geodesic equations given in eq.$\left(
\ref{6}\right) $, other than isometries.

Since the Noether symmetries form a subalgebra of the symmetries of
the geodesic equations, therefore, the Noether symmetries can not be
other than $h\oplus d_{2}$. As $h$ are the isometries,
$\mathbf{X}_{0}$ is associated with the Lagrangian, $L$, and $
\mathbf{X}_{1}$ is not a Noether symmetry by theorem 1. Therefore,
there does not exist new conservation law for the spaces of non-zero
curvature with no section of zero curvature. This completes the
proof of theorem 2.

\textbf{Example 1:} Consider Bertotti - Robinson like metric $\left[
4\right]$ (having 2-dimensional section of zero curvature)
\begin{equation}
ds^{2}=\cosh ^{2}\left( \frac{x}{a}\right) dt^{2}-dx^{2}-\left(
dy^{2}+dz^{2}\right) ,\label{9}
\end{equation}
and to determine Noether symmetries use
\[
L=\cosh ^{2}\left( \frac{x}{a}\right) \dot{t}^{2}-\dot{x}^{2}-\left(
\dot{y} ^{2}+\dot{z}^{2}\right) .
\]
This metric admits 9 Noether symmetries which are
\begin{eqnarray}
\mathbf{X}_{0} &=&\partial /\partial
t\,,\,\,\,\,\,\,\,\,\,\,\,\,\,\,\,\,\,\,\,\,\,\,\,\,\,\,\mathbf{X}
\,_{1}=\,z\partial /\partial y-y\partial /\partial z, \\
\mathbf{X}_{2} &=&\partial /\partial
z,\,\,\,\,\,\,\,\,\,\,\,\,\,\,\,\,\,\,\,\,\,\,\,\,\,\,\,\mathbf{X}
_{3}=\partial /\partial y, \\
\mathbf{X}_{4} &=&-\frac{1}{a}\tanh \frac{x}{a}\sin
\frac{t}{a}\partial /\partial t+\cos \frac{t}{a}\partial /\partial
x,\\
\mathbf{X}_{5} &=&\tanh \frac{x}{a}\cos \frac{t}{a}\partial
/\partial t+\sin
\frac{t}{a}\partial /\partial x, \\
\mathbf{X}_{6} &=&\partial /\partial s, \\
\mathbf{X}_{7} &=&s\frac{\partial }{\partial y}\text{\ \ \ \ with
gauge term
}A=2y \\
\mathbf{X}_{8} &=&s\frac{\partial }{\partial z}\text{\ \ \ \ with
gauge term }A=2z
\end{eqnarray}

The first six symmetries are the isometries of the spacetime and
$\mathbf{X}_{6} $ corresponds to the Lagrangian. The new conserved
quantities corresponding to $\mathbf{X}_{7}$ and $\mathbf{X}_{8}$
are $ T=s\dot{y}-y$ and $ T=s\dot{z}-z$ respectively.

\textbf{Example 2: }The Lagrangian for the line element (having zero
curvature everywhere)
\begin{equation}
ds^{2}=dx^{2}+ dy^{2}+dz^{2},
\end{equation}
is
\begin{equation}
L=\dot{x}^{2}+\dot{y} ^{2}+\dot{z}^{2} .
\end{equation}

The Noether symmetries are as follows
\begin{equation}
\mathbf{X}_{0}=\partial /\partial
x,\,\,\,\,\,\,\,\,\,\,\,\,\,\,\,\,\,\,\,\,\,\,\mathbf{X}_{1}=\partial
/\partial y,\,\,\,\,\,\,\,\,\,\,\,\,\,\,\,\,\,\,\,\,\,\,\mathbf{X}
_{2}=\partial /\partial z,
\end{equation}
\begin{eqnarray*}
\mathbf{X}_{3} &=&x\frac{\partial }{\partial z}-z\frac{\partial
}{\partial x}
, \\
\mathbf{X}_{4} &=&y\frac{\partial }{\partial x}-x\frac{\partial
}{\partial y}
, \\
\mathbf{X}_{5} &=&z\frac{\partial }{\partial y}-y\frac{\partial
}{\partial z}
, \\
\mathbf{X}_{6} &=&\frac{\partial }{\partial s}, \\
\mathbf{X}_{7} &=&s^{2}\frac{\partial }{\partial s}+sx\frac{\partial
}{
\partial x}+sy\frac{\partial }{\partial y}+sz\frac{\partial }{\partial z}
,\;\;\;\text{with gauge term \ }A=\left( x^{2}+y^{2}+z^{2}\right)  \\
\mathbf{X}_{8} &=&2s\frac{\partial }{\partial s}+x\frac{\partial
}{\partial x }+y\frac{\partial }{\partial y}+z\frac{\partial
}{\partial z},\;\;\;\text{
with gauge term \ }A=0 \\
\mathbf{X}_{9} &=&s\frac{\partial }{\partial x},\;\;\;\text{with
gauge term
\ }A=2x \\
\mathbf{X}_{10} &=&s\frac{\partial }{\partial y},\;\;\;\text{with
gauge term
\ }A=2y \\
\mathbf{X}_{11} &=&s\frac{\partial }{\partial z},\;\;\;\text{with
gauge term \ }A=2z
\end{eqnarray*}

The first six symmetries are the isometries of the space and
$\mathbf{X}_{6} $ corresponds to the Lagrangian. The new conserved
quantities corresponding to $\mathbf{X}_{7}$, $\mathbf{X}_{8}$,
$\mathbf{X}_{9}$, $\mathbf{X}_{10}$ and $\mathbf{X}_{11}$ are $
T=s^{2}L-2s(x\dot{x}+y\dot{y}+z\dot{z})+(x^{2}+y^{2}+z^{2})$, $
T=sL-(x\dot{x}+y\dot{y}+z\dot{z})$, $ T=s\dot{x}-x$, $ T=s\dot{y}-y$
and $ T=s\dot{z}-z$ respectively.

\textbf{Conclusion}

In this paper the new conserved quantities for the spaces of
different curvatures are discussed with the help of method
introduced by Lie. It is shown that for the spaces of non-zero
curvature there does not exist any conservation law other than the
Lagrangian and those associated with isometries. However, new
conserved quantities exist for the spaces of zero curvature or the
spaces having a section of zero curvature. It is worth mentioning
here that an n-dimensional space of zero curvature admits a total
number of $\frac{n(n-1)}{2}+2n+3=\frac{n^{2}+3n+6}{2}$ symmetries.
Among which $\frac{n(n-1)}{2}+n$ (rotation and translations) are the
isometries and one corresponds to the Lagrangian. And the remaining
$n+2$ correspond to the new conserved quantities. Also, for the
spaces with an m-dimensional section of zero curvature we have the
following

\textit{\textbf{Conjecture:} The spaces with an m-dimensional
section of zero curvature admit m new conserved quantities and the
corresponding Noether symmetries have the form
$s\frac{\partial}{\partial x^{i},}$ $i=1,2,...m$}.

\textbf{Acknowledgments}

The author is grateful to Professor Azad A. Siddiqui and Dr. Khalid
Saifullah for their valuable suggestions.

\end{document}